\documentclass[aps,prd,secnumarabic,amssymb, amsmath,nobibnotes,nofootinbib,11pt]{revtex4} 
\usepackage{amsfonts,amsmath,hyperref,url, color}
\usepackage{bm, bbm}
\usepackage{graphicx}
\usepackage{mathtools}

\newcommand{\be}{\begin{equation}}
\newcommand{\bal}{\begin{align}}
\newcommand{\eal}{\end{align}}
\newcommand{\ee}{\end{equation}}
\newcommand{\bea}{\begin{eqnarray}}
\newcommand{\eea}{\end{eqnarray}}
\newcommand{\bit}{\begin{itemize}}
\newcommand{\eit}{\end{itemize}}

\newcommand{\braket}[2]{\left\langle#1 |  #2\right\rangle}

\newcommand{\ket}[1]{\left|#1\right\rangle}

\newcommand{\ba}{\begin{aligned}}
\newcommand{\ea}{\end{aligned}}

\usepackage{color}

\begin{document}

\title{Vacuum thermal effects in flat space-time from conformal quantum mechanics}

\author{Michele Arzano}
\email{michele.arzano@na.infn.it}
\affiliation{Dipartimento di Fisica ``E. Pancini", Universit\`a di Napoli Federico II, I-80125 Napoli, Italy\\}
\affiliation{INFN, Sezione di Napoli,\\ Complesso Universitario di Monte S. Angelo,\\
Via Cintia Edificio 6, 80126 Napoli, Italy}

\begin{abstract}
The generators of radial conformal symmetries in Minkowski space-time can be mapped to the generators of time evolution in conformal quantum mechanics. Within this correspondence we show that in conformal quantum mechanics the state associated to the inertial vacuum in Minkowski space-time has the structure of a thermofield double. Such state is built from a bipartite ``vacuum state", the ground state of the generators of hyperbolic time evolution, which cover only part of the time domain. When time evolution is restricted to a finite time domain one obtains the temperature perceived by static diamond observers in the Minkowski vacuum. When time evolution is determined by dilations, covering only half of the time line, the temperature of the thermofield double corresponds to the non-vanishing temperature perceived by Milne observers whose proper time evolution is confined to the future cone (Milne universe) of Minkowski space-time. The two pictures are related by a conformal transformation on the real line. Our result provides a purely group theoretical derivation of the Milne and diamond temperature and shows that the fundamental ingredient for vacuum thermal effects is the presence of a horizon rather than acceleration.

\end{abstract}

\maketitle

\section{Introduction}
The notion of vacuum state in quantum field theory is observer dependent \cite{Davies:1984rk}. The importance of this basic fact was only started to be appreciated with the discovery of black hole quantum radiance \cite{Hawking:1974sw,Unruh:1976db}, which is one of its main consequences. The existence of a temperature and thermal features associated to the horizon of a black hole is directly related to two of the major open issues in quantum gravity: the fate of unitarity in the quantum evolution of black holes \cite{Hawking:1976ra,Mathur:2009hf,Giddings:2019vvj} and the nature of their Bekenstein-Hawking entropy \cite{Bekenstein:1973ur,Carlip:2008rk}, and it is at the basis of the interpretation of gravity as a ``thermodynamics of space-time" \cite{Jacobson:1995ab}.

The ambiguity in the choice of the vacuum state in curved space-time relies on the possibility of having different choices of time-like Killing vectors that can be used to decompose the space of solutions of the classical field equation into positive and negative energy subspaces \cite{Ashtekar:1975zn}. The former will define, together with a positive definite inner product, the one-particle Hilbert space for static observers with respect to the Killing vector which generates evolution in their proper time. From such Hilbert space the standard Fock space construction follows with the associated vacuum and multi-particle states \cite{Wald:1995yp}. The Hawking effect boils down to the fact that the Hartle-Hawking vacuum state associated to freely falling observer is a thermal state at the Hawking temperature for static observers with respect to Schwarzschild time whose associated Killing vector becomes null at the horizon.

In maximally symmetric spaces the quantization ambiguity associated to different possible choices of time evolution is eliminated by the requirement of invariance of the vacuum state under the full group of isometries. In Minkowski space-time this singles out the standard vacuum state associated to inertial time evolution under the ordinary time translation Killing vector. However, as it is well known, one can also consider time evolution generated by a boost Killing vector which is time-like in the Rindler wedges and whose orbits correspond to worldlines of uniformly accelerating observers. For such observers the inertial vacuum is populated by a thermal distribution of particles at a temperature proportional to the magnitude of their four-acceleration: the so-called Unruh temperature \cite{Unruh:1976db}. 
 
Quite interestingly, if one restricts the attention to scale-invariant fields the range of possible candidates for generators of time evolution can be extended to include {\it conformal Killing vectors}. In Minkowski space-time this has recently been used \cite{Higuchi:2017gcd,Wald:2019ygd} (see also \cite{Crispino:2007eb,Olson:2010jy}) to consider ``Milne quantization" of a two-dimensional massless scalar field in the future and past cones, analogous to Rindler quantization in the Rindler wedges, where time evolution is generated by the dilation conformal Killing vector. Observers along the radial orbits of dilations are comoving observers in a Milne universe (which is just Minkowski space-time described in expanding coordinates covering only the future cone) which perceive the inertial vacuum as populated by a thermal distribution of Milne particles.

These considerations suggest that in higher dimensions, and here we refer for simplicity to the $3+1$ dimensional case, one could define time evolution in terms of conformal Killing vectors which are time-like in certain regions of Minkowski space-time and similar thermal effects should manifest. Of particular interest are, to this extent, {\it radial conformal symmetries}, whose full classification for Minkowski space-time has been given in \cite{RCM}. Radial conformal Killing vectors divide Minkowski space-time into causal domains separated by light-like surfaces which are {\it conformal Killing horizons}, they are analogous to stationary spherically symmetric Killing horizons of black hole geometries and have similar thermodynamic properties \cite{DeLorenzo:2017tgx, DeLorenzo:2018ghq}.  Among such radial conformal Killing vectors are the ones which map a causal diamond into itself and whose orbits describe worldlines of observers confined within the diamond. The fact that these diamond observers should perceive the inertial Minkowski vacuum as a thermal state was first suggested in \cite{Martinetti:2002sz} where the claim was supported with arguments which used the so-called ``thermal time hypothesis''. Another attempt to provide evidence of such diamond temperature was given in \cite{Su:2016wop} using a model of Unruh-DeWitt detector moving along the diamond observer trajectory.

In \cite{Arzano:2020thh} the present author pointed out that radial conformal symmetries are in correspondence with generators of time evolution in conformal quantum mechanics, a $0+1$ dimensional conformal field theory \cite{deAlfaro:1976vlx}. Using this correspondence, for which the ``wordline quantum mechanics" of diamond observers at the origin\footnote{This result has analogies with the worldline quantum mechanics description of the temperature perceived by a static patch observer in de Sitter space \cite{Anninos:2011af,Nakayama:2011qh}} is viewed as a conformal quantum mechanics, it was shown that the Minkowski vacuum is perceived by diamond observers as a thermal state at a temperature inversely proportional to the size of the diamond. The results of \cite{Arzano:2020thh}, besides providing a simple but powerful tool for studying the thermodynamics of causal regions in Minkowski space-time, also show that the possibility of selecting different generators of time evolution in conformal quantum mechanics leads to thermal effects in complete analogy with higher dimensional quantum field theory. Since the focus of \cite{Arzano:2020thh} was on causal diamonds, and the corresponding hyperbolic time evolution restricted to a finite time domain in conformal quantum mechanics, the natural question is whether similar thermal effects can be associated to other choices of time evolution like, for example, the one determined by the generator of dilations, and what is the physical interpretation of the associated temperature in Minkowski space-time.

In this work we show that, besides the ``diamond temperature" associated to time evolution restricted to a finite time domain of conformal quantum mechanics, there is a natural temperature associated to time evolution generated by dilations and restricted to positive (or negative) times. These two thermal effects have their origin in the {\it thermofield double} structure of the vacuum state associated to the ``inertial" time evolution generator which is here evidenced using purely group theoretical arguments. The thermal effect associated to time evolution restricted to half of the time line indicates that in Minkowski space-time static Milne observers, i.e. observers at the origin whose proper time evolution is determined by dilations, perceive the Minkowski vacuum as a thermal state. Such result provides evidence for the existence of a ``Milne temperature" for space-time dimensions higher than two and exhibits its intimate connection with diamond temperature given that, as it shown below, the two pictures are connected on the conformal quantum mechanics side by a conformal transformation. Finally, since both the diamond and Milne temperatures are non-vanishing along the worldlines of static observers sitting at the origin with vanishing acceleration, our result demonstrates that the fundamental ingredient for quantum vacuum thermal effects is the presence of a horizon, in this case associated to the non-eternal nature of the lifetime of the observer, rather than a non-vanishing acceleration.

\section{Radial conformal Killing vectors in Minkowski space-time and time evolution in conformal quantum mechanics}

Let us start from the line element of Minkowski space-time written in spherical coordinates
\be\label{minksph}
d s^2 = -d t^2 + d r^2 + r^2 d \Omega^2\,,
\ee
where $d \Omega^2 = d \theta^2 + \sin^2 \theta\,  d\phi^2$. The most general radial conformal Killing vector $\xi$ for which $\mathcal{L}_{\xi} \eta_{\mu\nu} \propto \eta_{\mu\nu}$ where $\mathcal{L}_{\xi}$  is the Lie derivative and $\eta_{\mu\nu}$ is the Minkowski, metric is given by \cite{RCM}
\be\label{xi}
\xi = a K_0 + b D_0 + c P_0\,,
\ee
where $K_0$,  $D_0$ and $P_0$ generate, respectively, special conformal transformations, dilations and time translations and close the $\mathfrak{sl}(2,\mathbb{R})$ Lie algebra 
\be\label{sl1}
[P_0,D_0]= P_0\,,\qquad [K_0,D_0]= - K_0\,,\qquad [P_0,K_0]= 2 D_0\,.
\ee 
These generators can be written in terms of radial and time derivatives as
\bea\label{}
P_0 & = & \partial_t\\
D_0 & = & r\, \partial_r + t\, \partial_t\\
K_0 & = & 2 t r\, \partial_r + (t^2+r^2)\, \partial_t \,,
\eea
so that \eqref{xi} becomes
\be\label{xi2}
\xi = \left(a(t^2+r^2)+b t +c\right)\, \partial_t + r (2 a t + b)\, \partial_r\,,
\ee
and in light-cone coordinate $u=t+r$ and $v=t-r$ we have
\be
\xi = (a u^2 + b u + c)\, \partial_u +  (a v^2 + b v + c)\, \partial_v\,.
\ee
In conformal quantum mechanics the generators $K_0$,  $D_0$ and $P_0$  play the role of different time evolution operators. In terms of the {\it inertial} time variable $\tau$ associated to the generator $P_0$ they can be written as differential operators 
\bea\label{getau}
P_0 &=& \partial_{\tau}\\
D_0 &=& \tau\, \partial_{\tau}\label{dzee}\\
K_0 &=& \tau^2\, \partial_{\tau}\,.
\eea
The first important observation is that radial conformal Killing vectors for static observers at the origin $r=0$ and on the light-cones $u=const.$ or $v=const.$ coincide with the most general time evolution operator in conformal quantum mechanics
\be
G = a K_0 + b D_0 + c P_0 =  (a \tau^2 + b \tau + c)\, \partial_{\tau}\,.
\ee
The causal structure of the Killing vector \eqref{xi2} in Minkowski space-time has been studied in \cite{RCM}. The quantity $\Delta = b^2 - 4 ac$ characterizes the norm of the Killing vector and is used to classify different types of generators. The generator $P_0$  for which $\Delta < 0$  is everywhere time-like and in conformal quantum mechanics generates time evolution in terms of ordinary translation spanning the whole real (time) line. The generator $D_0$ with $\Delta > 0$ and $a=0$ is time-like inside the light-cone emanating from the origin, null on the light-cone and space-like outside. In conformal quantum mechanics it describes time evolution in terms of dilations restricted to the positive or negative real line. In Minkowski space-time it describes {\it conformal time} evolution in a Milne universe\footnote{The Milne universe being flat can be indeed mapped to the future (or past, inverting the time direction) inner light cone of Minkowski space-time. It corresponds to a slicing of the latter in terms constant time hyper-surfaces given by Euclidean hyperboloids, orbits of boosts.} \cite{RCM,Olson:2010jy,Wald:2019ygd}. The other relevant generator is the radial conformal Killing vector which maps a causal diamond, i.e. the region $|t|+|r|<\alpha$, into itself given by
\be\label{zeta}
S_0 =  \frac{1}{2} \left(\alpha P_0 - \frac{K_0}{\alpha}\right) = \frac{1}{2 \alpha} \left((\alpha^2-t^2-r^2) \partial_t - 2 t r \partial_r \right)\,.
\ee
The causal structure of this Killing vector for which $\Delta > 1$ and $a \neq 0$ is more articulated. It is null on the light-cones emanating from $r=0$ at times $\alpha$ and $-\alpha$. It is time-like inside or outside both light-cones, i.e. within the {\it causal diamond} of radius $\alpha$ and its causal complement, and space-like everywhere else. In conformal quantum mechanics the combination $S_0$ generates time evolution restricted to the finite domain $t\in (-\alpha, \alpha)$, the analogue of the {\it diamond} on the real line.

Notice that the quantity $\Delta = b^2 - 4 ac$ is proportional to the determinant of $G$ seen as a $2\times 2$ matrix and thus it is invariant under the adjoint action of $SL(2,\mathbb{R})$ on its Lie algebra. Such determinant can be used to classify different types of transformations generated by $G$. In particular while $P_0$, for which $\Delta < 0$, generates {\it parabolic} transformations of the Euclidean plane (null rotations if we think of the relation between $SL(2,\mathbb{R})$ and the three dimensional Lorentz group $SO(2,1)$), $D_0$ and $S_0$, for which $\Delta > 0$, generate $SL(2,\mathbb{R})$ elements belonging to the class of {\it hyperbolic} transformations whose representatives on the Lorentz group side are boost transformations. Thus we can map one into another using an $SL(2,\mathbb{R})$ transformation.

In order to do so let us first introduce time coordinates adapted to the evolution defined by the generators $D_0$ and $S_0$ i.e. such that 
\bea
\frac{D_0}{\alpha}  &= &\partial_{\nu}\label{adnusig}\\
\frac{S_0}{\alpha}  &=& \partial_{\sigma}\label{adnusig2}\,,
\eea
defining the Milne and diamond time for static observers with semi-infinite and finite lifetimes respectively. We can easily derive the transformation between the inertial time $\tau$ and the Milne and diamond time coordinates $\nu$ and $\sigma$ comparing \eqref{adnusig} and \eqref{adnusig2} with the expressions \eqref{dzee} and \eqref{zeta}. One obtains
\bea
\nu & = & \alpha \log{|\tau|} + C_1\\
\sigma & = & \alpha \log{\frac{|\tau+\alpha|}{|\tau-\alpha|}} + C_2\,.
\eea
where the integration constants $C_1$ and $C_2$ can be determined by requiring dimensional consistency. We set
\bea
C_1 & = & - \log{2\alpha}\\
C_2 & = & 0\,,
\eea
the reason for the choice $2 \alpha$ for the argument of the $\log$ will be clear later. We thus have for the Milne time 
\bea
\tau = &2\alpha\, \exp{\frac{\nu}{\alpha}}& \qquad \text{for} \,\, \tau>0\label{ttaup}\\
\tau = & - 2\alpha\, \exp{\frac{\nu}{\alpha}}& \qquad \text{for} \,\, t<0\,.
\eea
We see that on the positive half-line $\tau>0$ as $\nu$ ranges from $-\infty$ to $\infty$ we go from $\tau=0$ to $\tau=\infty$. On the negative half line the time flow is inverted and we go from $\tau=0$ to $\tau=-\infty$ as $\nu$ ranges from $-\infty$ to $\infty$. Notice that, even though there is no natural time scale associated with the future cone, a constant with dimensions of time is needed in order to define a time evolution operator in terms of the generator of dilations $\tau\, \partial_\tau$. This time scale is the point on the real line measured in time $\tau$ corresponding to the origin $\nu=0$ of Milne time.\\
For the $S$-time variable $\sigma$ we have that in the region $|\tau|<\alpha$, i.e. within the diamond, i.e.
\be
\tau = \alpha\, \tanh{\sigma/2\alpha}
\ee
while for $|\tau| > \alpha$ one has
\be
\tau = \alpha \coth{\sigma/2\alpha}\,.
\ee
We see that within the diamond as the diamond time ranges from $-\infty$ to $\infty$ we cover the region from $t=\alpha$ to $t=-\alpha$.

Let us now consider a new time coordinate $\tau'(\tau)$ in terms of which the dilation generator $D = \tau\, \partial_\tau$ acts as the generator $S$ of conformal transformations which preserve the diamond, i.e.  
\be
 \tau\, \partial_\tau = \frac{1}{2 \alpha} \left(\alpha^2- \tau'^2\right) \partial_{\tau'}\,.
\ee 
Such equation can be easily solved leading to 
\be
\log{\frac{|\tau'+\alpha|}{|\tau'- \alpha|}} = \log{|\tau|} + C \,. 
\ee
For dimensional consistency we choose for the integration constant $C= -\log{A}$, where $A$ has dimensions of time and thus must be proportional to $\alpha$. Restricting to the positive half-line $\tau > 0$ we have
\be\label{ttpa}
\tau = A\,\, \frac{\tau'+\alpha}{\tau'-\alpha}
\ee
In order to determine the constant $A$ we first notice that the point $\tau'=-\alpha$ gets mapped to $\tau=0$ while $\tau=\alpha$ to $\tau = \infty$. The origin of the diamond $\tau'=0$ gets mapped to $\tau= -A$ so we know that $A$ must be negative. Let us further notice that the origin of the diamond proper time $\sigma=0$ corresponds to $\tau'=0$. Under the map \eqref{ttpa} this is mapped to $\tau=-A$. Now we require that the origins of the diamond and Milne proper times are ``syncronized" i.e. the point corresponding to $\sigma=0$ in $\tau'$ coordinates is mapped to the point corresponding to $\nu=0$ in $\tau$-coordinates and thus, from \eqref{ttaup},
\be
2 \alpha = -A\,.
\ee
We thus have
\be\label{DtoS}
\tau = - 2 \alpha\, \frac{\tau'+\alpha}{\tau'-\alpha}
\ee
which can be inverted to give
\be\label{StoD}
\tau' =  \alpha\, \frac{\tau - 2\alpha}{\tau + 2 \alpha}\,.
\ee
Notice how this transformations correspond to the map from the causal diamond to the Rindler wedge and viceversa \cite{Casini:2011kv, Jacobson:2018ahi} when written in radial light-cone coordinates, indeed the Rindler wedge and the future cone meet on the light cone.

\section{The thermofield double of conformal quantum mechanics}
As shown in \cite{Chamon:2011xk,Jackiw:2012ur} conformal quantum mechanics can be seen as a one-dimensional conformal field theory in which correlation functions are built from eigenstates of the time translation operator. Let us recall the main features of this construction. In conformal quantum mechanics we work with the ``quantum mechanical" counterparts of the generators introduced in the previous section i.e. $H = iP_0$, $D= iD_0$ and $K=i K_0$ closing the algebra
\be\label{sl2}
[H,D]=i H\,,\qquad [K,D]= - i K\,,\qquad [H,K] = 2 i D\,.
\ee
Let us look for states $|\tau \rangle$ on which the $SL(2, \mathbb{R})$ generator $H$ acts as a $\tau$-derivative
\be
H |\tau \rangle = -i\, \partial_\tau\,  |\tau \rangle\,. 
\ee
These states were first constructed in \cite{deAlfaro:1976vlx} starting from an irreducible representation of the $\mathfrak{sl}(2,\mathbb{R})$ Lie algebra. Upon defining the operators 
\be
L_\pm  \equiv  \frac{1}{2}\, \left(\frac{K}{\alpha} - \alpha\, H \right)\, \pm i\, D,  
\ee
and
\be
L_0  \equiv  \frac{1}{2}\, \left(\frac{K}{\alpha} + \alpha\, H \right)\,,
\ee
with commutation relations
\begin{equation}\label{eq:su11Alg}
[L_-,L_+] = 2 L_0\,, \quad [L_0, L_{\pm}] = \pm L_{\pm}\,,
\end{equation}
such representation is realized in terms of the eigenstates $\ket{n}$ of $L_0$ 
\begin{subequations}\label{rjAdS6}
\begin{gather}
L_0 \ket{n} =  r_n\, \ket{n}\label{rjAds6-a}\\
r_n = r_0 + n,\ \  r_0 > 0, \ \ n = 0,1 \ldots \nonumber\\
\braket{n}{n^\prime} = \delta_{n\, n^\prime} \nonumber \\
L_\pm \ket{n} = \sqrt{r_n \, (r_n \pm 1) - r_0 \, (r_0 -1)} \, \ket{n \pm 1} \label{rjAds6-b}
\end{gather}
\end{subequations}
where $r_0$ is related to the eigenvalue of the Casimir operator of the $\mathfrak{sl}(2,\mathbb{R})$ algebra
\be
\mathcal{C}= \frac{1}{2}\left(K H + H K \right) - D^2 = r_0 (r_0-1)\,.
\ee
Notice that dimensional consistency requires the introduction of a scale $\alpha$ with the dimension of time in the definition of the operators $L_+$ and $L_-$ and, in light of the correspondence with conformal symmetries in Minkowski space-time, we can give a geometric interpretation to such scale, following the discussion in the previous section, in terms of the radius of a causal diamond whose boundaries are invariant under the flow of the conformal Killing vector $S_0$.

In \cite{Chamon:2011xk,Jackiw:2012ur} it was found that the $|\tau \rangle$ states can be obtained from the $ \ket{n=0} $ ``vacuum state" via
\be\label{opcft}
|\tau \rangle = \mathcal{O}(\tau) \ket{n=0}  = N(\tau)\exp(-\omega(\tau)L_+)  \ket{n=0}\,,
\ee
where
\begin{eqnarray}
N(\tau) &=& [\Gamma(2r_0)]^{1/2}\left(\frac{\omega(\tau)+1}{2}\right)^{2r_0} \ , \nonumber \\
\omega(\tau) &=& \frac{a+i\tau}{a-i\tau} \ .
\end{eqnarray}
The action of the other generators on such $\tau$-states are given by
\bea
D \ket{\tau} &=& - i\, \left(\tau\, \frac{d}{d \tau} + r_0 \right)  \ket{\tau},\\
K \ket{\tau} &=& -i \left(\tau^2\,  \frac{d}{d \tau} + 2\, r_0\, \tau \right) \ket{\tau}\,.
\eea
In \cite{Chamon:2011xk}  the two point functions of the theory are identified with the inner product between the $\tau$-states
\be\label{2pf}
G_2 (\tau_1, \tau_2) \equiv \braket{\tau_1}{\tau_2} = \frac{\Gamma\, (2 r_0)\, \alpha^{2 r_0}}{[2 i\, (\tau_1 - \tau_2)]^{2 r_0}}\,,
\ee
where $r_0$ plays the role of the conformal weight. General $n$-point functions with $n>2$ are obtained by inserting generic primary operators between $\langle \tau_1|$ and $|\tau_2\rangle$. When $\tau=0$ one has $\omega(\tau) = 1$ and thus 
\be\label{tze1}
|\tau =0 \rangle = \Gamma(2r_0)^{1/2}\, \exp(-L_+)  \ket{n=0} 
\ee
One immediately sees from \eqref{2pf} that such $\tau$-vacuum is not normalizable due to the divergence of the two-point function for coincident points.

In \cite{Arzano:2020thh} it was noted that for $r_0 = 1$ the two-point function \eqref{2pf} is proportional to the two-point function of a free massless scalar field in Minkowski space-time evaluated along the trajectory of a static inertial observer at the origin. This fact was used to derive the diamond temperature\footnote{It should be noted that $H$ and $D$ form a sub-algebra of $\mathfrak{sl}(2,\mathbb{R})$ which generates the group of affine transformations of the real line, also known as the $ax+b$ group. In \cite{Arzano:2018oby} it was shown, using one-dimensional fields carrying a representation of the $ax+b$ group and Bogoliubov transformations between $H$-modes and $D$-modes, that the $H$-vacuum is seen as a thermal distribution of $D$-particles when restricted to the positive half line.} upon re-writing \eqref{2pf} in terms of the time variable associated to the diamond time evolution generator $S=iS_0$.  Here we provide an alternative derivation of the same temperature by looking directly at the structure of the $\tau$-vacuum. This will lead to a unified description of the diamond and Milne temperatures.

Let us start from a simple observation which, nonetheless, is a key passage of our derivation. It is known from the quantum optics literature (see e.g. \cite{Ban:1992zz,Chaturvedi:1991zzb}) that the operators $L_{\pm}$ and $L_0$ with commutation rules \eqref{eq:su11Alg} can be realized in terms of {\it two sets} of creation and annihilation operators $a_L, a_L^\dagger, a_R,  a_R^\dagger$ as
\begin{equation}\label{casl}
L_+=a_L^\dagger a_R^\dagger\,, \quad L_- = a_L a_R\,,
\quad L_0=\frac{1}{2}\left( a_L^\dagger a_L + a_R^\dagger a_R+1\right)\,.
\end{equation}
It immediately follows from \eqref{tze1} that the $\tau$-vacuum can be written as\footnote{From this point on we set $r_0=1$ as in \cite{Arzano:2020thh}.}
\be\label{lasl}
|\tau =0 \rangle  =  \exp\left[ - a^\dagger_L a^\dagger_R\right]|0\rangle_L  \otimes  |0\rangle_R~,
\ee
where we made explicit the bipartite structure of the vacuum state $\ket{n=0}$ emerging from the decomposition \eqref{lasl} 
\be
\ket{n=0} = |0\rangle_L  \otimes  |0\rangle_R\,.
\ee
With simple manipulations we have that 
\be\label{ttens}
\ba
|t =0 \rangle = &   \sum_{n=0}^\infty \frac{(-1)^n}{n!}\,\left(a^\dagger_L a^\dagger_R\right)^n|0\rangle_L \otimes |0\rangle_R\\
= &  \, \sum_{n=0}^\infty (-1)^n  |n\rangle_L \otimes  |n\rangle_R\,,
\ea
\ee
where $|n\rangle_L$ and $|n\rangle_R$ are eigenstates of the ``left" and ``right" number operators $N_L= a^\dagger_L a_L $ and $N_R= a^\dagger_R a_R$. Our second key observation is that the Lie algebra 
\begin{equation}\label{lzeropm}
[L_-,L_+] = 2 L_0\,, \quad [L_0, L_{\pm}] = \pm L_{\pm}\,,
\end{equation}
can be realized via another combination of $H$, $D$ and $K$, namely 
\be\label{lzeros}
L_0= i S\,,\qquad L_+= \frac{1}{2}\left(D-R\right)\,,\qquad L_-=2 \left(D+R\right)\,,
\ee
where $R=\frac{1}{2}\, \left(\frac{K}{\alpha} + \alpha\, H \right)$ is generator of elliptic transformations corresponding to rotations in the three-dimensional Lorentz group. Under the identification \eqref{lzeros} we can now interpret the $|n\rangle$ states as eigenstates of the diamond time evolution generator $S$ and, in particular, the bipartite vacuum $\ket{n=0} = |0\rangle_L  \otimes  |0\rangle_R$ as the ``ground state'' of such Hamiltonian. In light of the correspondence with radial conformal symmetries in Minkowski space-time, we can think of the vacuum state $\ket{n=0}$ as the conformal quantum mechanics counterpart of the the vacuum state associated the generator of time evolution for static diamond observers, i.e. as the diamond analogue of the Boulware vacuum for Rindler observers.

Along the same lines we can think of the $\tau$-vacuum $|\tau =0 \rangle$ as the conformal quantum mechanics counterpart of the Hartle-Hawking vacuum of a massless scalar field in Minkowski space-time. Such connection is a bit more subtle and should be understood in terms of a state-operator correspondence illustrated in \cite{Chamon:2011xk}. As observed in \cite{Arzano:2020thh} the Minkowski light-cone two-point function maps to the $CFT_1$ two-point function constructed from the $|\tau \rangle$ states \eqref{2pf}. The latter can be written as 
\be
\langle \tau_1|\tau_2\rangle = \langle n=0 |\, \mathcal{O}(\tau_1)\, \mathcal{O}(\tau_2)\,| n=0 \rangle\,,
\ee
with $\mathcal{O}(\tau)$ defined in \eqref{opcft}. The vacuum sate $| n=0 \rangle$ is not conformally invariant and the operators $\mathcal{O}(\tau)$ do not transform as primary operators under conformal transformations \cite{Chamon:2011xk}. However, as shown in \cite{Chamon:2011xk}, the combination
\be
|\tau\rangle = \mathcal{O}(\tau)| n=0 \rangle 
\ee
satisfies a constraint which combines both the requests of the invariance of the vacuum with that of the operators $\mathcal{O}_{r_0}(t)$ to be primary. In other words the two point functions $\langle t_1|t_2\rangle$ behave as if they were constructed from a conformal invariant vacuum in analogy with the two-point function for massless scalar field in Minkowski space-time constructed from the Hartle-Hawking vacuum. Indeed it turns out \cite{Chamon:2011xk} that such two-point functions can be expressed as
\be
\langle \tau_1|\tau_2\rangle = \langle \tau=0 | e^{-i (\tau_1 - \tau_2)H}| \tau=0 \rangle\,,
\ee
i.e. a quantum mechanical counterpart of the Minkowski space-time two-point function with $| \tau=0 \rangle$ playing the role of the Hartle-Hawking state.

The analogy between the $\tau$-vacuum and the Hartle-Hawking state is actually deeper than it what might appear at first glance. Indeed the state $| \tau=0 \rangle$, like the Hartle-Hawking vacuum, exhibits the structure of a {\it thermofield double}. To see this let us recall the expression of $L_0$ in terms of creation and annihilation operators \eqref{casl}. Using the identification \eqref{lzeros}, from \eqref{ttens} we can write the $\tau$-vacuum as 
\be\label{tfdc1}
|\tau =0 \rangle = -i \, \sum_{n=0}^\infty  e^{i\pi L_0}\,  |n\rangle_L|n\rangle_R\, = -i \, \sum_{n=0}^\infty  e^{-\pi S}\,  |n\rangle_L|n\rangle_R\,.
\ee
Such superposition of states is the same as the one appearing in the thermofield double state of a bosonic oscillator \cite{Khanna:2009zz,Chapman:2018hou}. Indeed, starting from the $\tau$-vacuum density matrix $\rho = |\tau =0 \rangle \langle \tau=0|$ and tracing over one set of $L$ or $R$ degrees of freedom one obtains a thermal density matrix at temperature $1/2\pi$ with respect to the modular Hamiltonian $S$. Now, of course, $S$ does not have the dimensions of a Hamiltonian, rather one should see the operator $\frac{S}{\alpha}$ as the modular Hamiltonian of the system so that the temperature associated to the thermofield double \eqref{tfdc1} is given by
\be
T_{S} = \frac{1}{2\pi \alpha}\,,
\ee
i.e. precisely the diamond temperature obtained in \cite{Arzano:2020thh}.

An interesting upshot of the analysis above is that it can be used to also give a group theoretical derivation of the Milne temperature. Indeed, under the conformal map \eqref{StoD} the identification \eqref{lzeros} leads to the following realization of the algebra \eqref{lzeropm}
\be\label{lzerod}
L_0= i D\,,\qquad L_+= -\alpha H\,,\qquad L_-=\frac{K}{\alpha}\,.
\ee
With such identification the vacuum state $\ket{n=0}$ is now seen as the conformal quantum mechanics analogue of the vacuum state associated to $D$, the generator of conformal time translations in the future cone, and the temperature 
\be
T_{D} = \frac{1}{2\pi \alpha}\,,
\ee
is simply the temperature perceived by a Milne observer sitting at the origin of Minkowski space-time in the inertial vacuum. In two space-time dimensions, where null planes and null cones coincide and Milne and Rindler particles can de identified on the the light cone \cite{Wald:2019ygd}, our arguments give a unified, group theoretical derivation of the diamond, Milne and Unruh temperatures.

\section{Conclusions}
The considerations above show that the structure of conformal quantum mechanics is rich enough to reproduce, in the extremely simplified context of a one-dimensional model, the basic features which characterize vacuum thermal effects in quantum field theory, i.e. the appearance of the global vacuum as a thermal state populated of excitations associated to Hamiltonians whose evolution does not cover the entire time domain of the theory. Remarkably, using the correspondence between time evolution in conformal quantum mechanics and radial conformal symmetries in Minkowski space-time seen as generators of proper time, this result provides direct evidence for the fact that the inertial vacuum appears as a thermal state to observer whose time evolution is not {\it eternal}. The results presented suggest that the thermodynamic properties of causal diamonds and the Milne ``patch" of Minkowski space are intimately connected. It is tempting to speculate that the tools of conformal quantum mechanics which are used to provide evidence for such connection might provide an alternative and perhaps simpler way to describe entanglement features of quantum fields across the causal boundaries of such regions. A thorough exploration of such hypothesis is left to future studies.


\end{document}